\documentclass[
	aps, prd, reprint,
	10pt, notitlepage, a4paper,
        floats, floatfix,
	amsmath, amssymb, amsfonts, 
	superscriptaddress,
	showpacs, showkeys,
	nofootinbib,
]{revtex4-1}

\def\paper{paper}
\newcommand{\append}[1]{{#1}}

\newcommand{\remark}[1]{{\color{red}#1}}
\newcommand{\longremark}[1]{{\color{red}$\blacksquare$}\footnote{\color{red}#1}}
\renewcommand{\remark}[1]{}
\renewcommand{\longremark}[1]{}

\usepackage{ifthen}
\newboolean{prd}
\setboolean{prd}{false}
\newboolean{arxiv}
\setboolean{arxiv}{true}

\newboolean{notprd}
\setboolean{notprd}{true}
\ifprd
\setboolean{notprd}{false}
\fi
\newboolean{notarxiv}
\setboolean{notarxiv}{true}
\ifarxiv
\setboolean{notarxiv}{false}
\fi

\usepackage{graphicx}

\usepackage[usenames]{xcolor}

\ifnotprd
\xdefinecolor{mylinkcolor}{rgb}{0,0,0.5}
\usepackage[
	bookmarksnumbered, bookmarksopen, bookmarksopenlevel=2,
	breaklinks=true,
	colorlinks=true, filecolor=mylinkcolor, citecolor=mylinkcolor,
	linkcolor=mylinkcolor, urlcolor=mylinkcolor, menucolor=mylinkcolor,
]{hyperref}

\def\nl{\\ & \quad}

\allowdisplaybreaks

\begin{document}

\ifnotprd
\hypersetup{
	pdftitle={Spin gauge symmetry in the action principle for classical relativistic particles},
	pdfauthor={Jan Steinhoff}
}
\fi

\title{Spin gauge symmetry in the action principle for classical relativistic particles}

\author{Jan Steinhoff}
\email{jan.steinhoff@aei.mpg.de}
\homepage{http://jan-steinhoff.de/physics/}
\affiliation{Max-Planck-Institute for Gravitational Physics (Albert-Einstein-Institute),
        Am M{\"u}hlenberg 1, 14476 Potsdam-Golm, Germany, EU}
\affiliation{Centro Multidisciplinar de Astrof\'isica --- CENTRA, Departamento de F\'isica,
	Instituto Superior T\'ecnico --- IST, Universidade de Lisboa --- ULisboa,
	Avenida Rovisco Pais 1, 1049-001 Lisboa, Portugal, EU}

\date{\today}

\begin{abstract}
We suggest that the physically irrelevant choice of a representative worldline of a
relativistic spinning particle should correspond to a gauge symmetry in an action
approach. Using a canonical formalism in special relativity, we identify a (first-class)
spin gauge constraint, which generates a shift of the worldline together with the corresponding
transformation of the spin on phase space. An action principle is formulated for which a minimal
coupling to fields is straightforward.
The electromagnetic interaction of a monopole-dipole particle is constructed explicitly.
\end{abstract}


\maketitle

\section{Introduction}
Relativistic spinning particles are an important topic in both
classical and quantum physics. All experimentally verified elementary particles,
except the Higgs boson, are spinning. In the classical regime, spinning
particles in relativity are a seminal topic, too, see \cite{Fleming:1965,Hanson:Regge:1974} for reviews.
But we are going to argue in this {\paper} that, as of now,
their formulation through a classical action principle remains incomplete.

\begin{figure}
\includegraphics{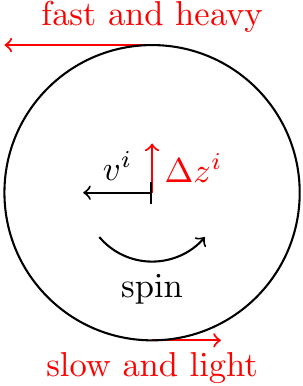}
\caption{If a spinning spherical symmetric object moves with a velocity $v^i$ to the left,
then its upper hemisphere moves faster with respect to the observer
than its lower hemisphere. Hence, the upper hemisphere possesses a larger relativistic
mass than the lower one and the object acquires a mass dipole $S^{i0} = m \Delta z^i$.
See, e.g., \cite{Fleming:1965}.\label{centerfig}}
\end{figure}
An interesting, but sometimes problematic, feature of spinning objects
in relativity is that their center of mass depends on the observer. This is vividly illustrated
in Fig.~\ref{centerfig}. On the other hand, the definition of the angular momentum of
the object, or spin, hinges on the location of the center as the reference point. Hence, both the three-dimensional
center and the three-dimensional spin of an object depend on the observer.
However, both are combined to form an antisymmetric four-dimensional spin-tensor
$S^{\mu\nu}$, where the spatial components $S^{ij}$ contain the usual spin
(or flux dipole) and the components $S^{i0}$ give the mass dipole.
The latter is non-zero if and only if the observed center of mass
differs from the reference point.

Actually, in special relativity, $S^{i0}$ is in one-to-one correspondence
with the choice of a reference point. This allows one to
represent the choice of center through a condition on $S^{\mu\nu}$,
called spin supplementary condition (SSC).
For instance, the physical meaning of the Fokker-Tulczyjew condition $S_{\mu\nu} p^{\nu} = 0$
\cite{Fokker:1929, Tulczyjew:1959} is that
one observes a vanishing mass dipole $S^{i0} = 0$
in the rest-frame of the object, where its momentum $p^{\nu}$ is parallel to
the time direction.
In other words, the chosen center agrees with the center of mass observed in the
rest-frame.
Because physically the dynamics of the object must be independent of the reference
worldline $z^{\mu}(\sigma)$
we chose to represent its motion, this choice is a gauge choice and the
SSC can be understood as a gauge fixing. However, we will see that, in the context
of an action formulation for spinning particles, there is a twist in this story.
In general relativity, the correspondence between the choice
of center and the SSC was rigorously demonstrated only for the
special case $S_{\mu\nu} p^{\nu} = 0$ \cite{Schattner:1979:1,Schattner:1979:2},
but intuitively one should expect that it holds more generally, too.

To date, the prototype construction for an action and canonical formalism
of classical spinning particles in special relativity is given in a
seminal paper of Hanson and Regge \cite{Hanson:Regge:1974}.
(For earlier work on the action see \cite{Goenner:Westpfahl:1967, Westpfahl:1969:2} and for
the canonical formalism see \cite{Kunzle:1972}). However, this
construction essentially covers the choice $S_{\mu\nu} p^{\nu} = 0$ only,
which has the advantage of being a covariant condition. Then the
covariance of the theory is manifest. But following the idea
that the choice of a reference worldline can be seen as an arbitrary gauge choice,
one would instead expect that the action possesses a corresponding gauge symmetry,
so that any SSC can be used equivalently at the level of the action.
It is the main purpose of the present {\paper} to work out a recipe for
the construction of such a gauge-invariant action in special relativity,
and in this sense completes the work in \cite{Hanson:Regge:1974}.

It is important to notice that Ref.~\cite{Hanson:Regge:1974} is the basis
of current methods for the computation of general relativistic spin effects
in compact binaries \cite{Porto:2006, Porto:Rothstein:2006, Porto:Rothstein:2008:2,
Levi:2008, Steinhoff:Schafer:2009:2, Steinhoff:2011, Levi:2010, Buonanno:Faye:Hinderer:2013, Blanchet:Buonanno:LeTiec:2013, Marsat:2014, Blanchet:2014}. Predictions
for these spin effects are of great importance for future gravitational wave
astronomy. For instance, some black holes are known to spin rapidly
\cite{Reynolds:2013, McClintock:Narayan:Steiner:2013}
and certain spin orientations lead to an increased gravitational wave luminosity
\cite{Reisswig:Husa:Rezzolla:Dorband:Pollney:Seiler:2009}.
This makes it likely that spin effects are relevant for the first detectable
sources.
If spinning compact objects are modeled through an effective
field theory approach (EFT) \cite{Porto:2006, Levi:2010, Goldberger:Rothstein:2006},
then it is vital to implement all expected symmetries in the effective action.
However, the gauge symmetry related to the choice of a representative worldline
was not considered so far.

It is noteworthy that classical spinning particles are based on a spin-1
representation in the present {\paper}, while Fermions transform under fractional spin representations.
But still, an action for classical spinning particles can also be seen as an effective
theory for fermion fields in certain limits. The clearest way to understand 
the classical limit is through a Foldy-Wouthuysen transformation \cite{Foldy:Wouthuysen:1950, Silenko:2003}.
This is a unitary transformation of the spinors, which removes the Zitterbewegung.
This unitary representation corresponds to the choice of the Pryce-Newton-Wigner center
\cite{Pryce:1935, Pryce:1948, Newton:Wigner:1949}
for the representative point in the classical theory. Then the commutators, or Poisson
brackets, of the three-dimensional spin and position are the (rather simple) standard
canonical ones.

We adopt the conventions from \cite{Hanson:Regge:1974}.

\section{Spinning particles}
Before we discuss classical spinning particles, it is useful to
recapitulate the definition of particles in quantum theory.
One-particle states are defined by the property of transforming
under an irreducible representation of the Poincar{\'e} group \cite{Weinberg:1995}.
That is, a particle is still the same one if it is transformed to a different
position, orientation, or speed by the Poincar{\'e} group and the irreducibility
guarantees that it can not be separated into parts, so it is indeed just one particle.
The linear momentum of the particle $p_{\mu}$ follows as the eigenvalue of the
translation operator. Since $p_{\mu}$ transforms as a vector
under Lorentz transformations, it can not label the irreducible
representations, but the scalar $p_{\mu} p^{\mu} = \mathcal{M}^2$ can.
Now, all vectors on such a mass shell are connected by Lorentz
transformations, so one can define a standard Lorentz transformation
$L(p)$ which brings $p_{\mu}$ to some standard form. A generic
Lorentz transformation of the particle state can then be decomposed
into the standard Lorentz transformation $L(p)$ and an element of the so called
little group. The latter leaves the standard form of $p_{\mu}$ invariant.
For instance, for massive particles
($\mathcal{M}^2>0$) one usually chooses the standard form of $p_{\mu}$
to point to the time direction ($p_{\mu}$ is boosted to the rest frame)
and the little group is the rotation group SO(3) (Wigner rotation),
since it leaves the time direction invariant. Finally,
this simple group-theoretical definition of particles implies that particles
are characterized by a mass shell and an irreducible representation
of the little group, which carries a spin quantum number.
An analogous group-theoretical approach can be used in the classical context \cite{Bel:Martin:1980}.

For classical particles, we follow \cite{Hanson:Regge:1974} and
represent the configuration space of the particle by a position $z^{\mu}$ and a Lorentz
matrix $\Lambda_A{}^{\mu}$, with obvious transformation properties
under the Poincar{\'e} group. Here the index $A$ labels the basis of the
body-fixed frame. Though this is borrowing terminology from a rigid body,
we insist that the constructed action can serve as an effective description
of generic spinning bodies. The generators of the Poincar{\'e} group are the
linear momentum $p_{\mu}$ and the total angular momentum $J_{\mu\nu} = 2 z_{[\mu} p_{\nu]} + S_{\mu\nu}$, so the Poisson brackets must read
\begin{align}\label{poisson}
\{ z^{\mu}, p_{\nu} \} &= \delta^{\mu}_{\nu} , \quad
\{ \Lambda_A{}^{\alpha}, S_{\mu\nu}\} = 2 \Lambda_A{}^{\beta} \eta_{\beta[\mu} \delta^{\alpha}_{\nu]} , \\
\{ S_{\alpha\beta}, S_{\mu\nu} \} &= S_{\alpha\mu} \eta_{\nu\beta} - S_{\alpha\nu} \eta_{\mu\beta} + S_{\beta\nu} \eta_{\mu\alpha} - S_{\beta\mu} \eta_{\nu\alpha} , \nonumber
\end{align}
all other zero, where $\eta_{\mu\nu}$ is the Minkowski metric.

We consider the case of a massive particle here. Then we
can define a standard boost which transforms the time direction
of the body-fixed frame $\Lambda_0{}^{\mu}$ into the direction of the
linear momentum $p_{\mu}$,
\begin{equation}\label{standardboost}
L^{\mu}{}_{\nu} = \delta^{\mu}{}_{\nu} + 2 \frac{p^{\mu} \Lambda_{0\nu}}{p}
	- \frac{p \omega^{\mu} \omega_{\nu}}{p_{\rho} \omega^{\rho}} ,
\end{equation}
where $p:=\sqrt{p_{\mu}p^{\mu}}$ and for later convenience we have introduced
the time-like vector $\omega^{\nu} := p^{\nu} / p + \Lambda_{0}{}^{\nu}$.
When applied to the Lorentz matrix, we obtain a new matrix
$\tilde{\Lambda}_A{}^{\mu} = L^{\mu}{}_{\nu} \Lambda_A{}^{\nu}$, or
explicitly
\begin{equation}\label{linv}
\tilde{\Lambda}_0{}^{\mu} = \frac{p^{\mu}}{p} , \quad
\tilde{\Lambda}_i{}^{\mu} = \Lambda_i{}^{\mu} - \Lambda_i{}^{\nu}
	\frac{p_{\nu} \omega^{\mu}}{p_{\rho} \omega^{\rho}}.
\end{equation}
We notice that $\tilde{\Lambda}_0{}^{\mu}$ is redundant, since it is
given by $p_{\mu}$. The independent physical degrees of freedom are contained in 
$\tilde{\Lambda}_i{}^{\mu}$, which carries an SO(3) index and hence
transforms under the vector (spin-1) representation of the little group.
This is intuitively clear, since the state of motion is already
characterized by the linear momentum and the temporal component of the
body-fixed frame must be redundant. The physical information of the body-fixed
frame is the orientation of the object, which should be associated with
three-dimensional rotations.

Because $\Lambda_A{}^{\mu}$ contains irrelevant, or gauge, degrees of freedom,
its conjugate $S_{\mu\nu}$ must be subject to a constraint. This constraint is usually
also the generator of the gauge symmetry \cite{Pons:2005}, which must leave the physical degrees
of freedom $\tilde{\Lambda}_i{}^{\mu}$ invariant. From the projector-like structure of Eq.~(\ref{linv})
and the fact that the spin generates Lorentz transformations of $\Lambda_i{}^{\mu}$,
we may guess that the generator is given by $S_{\mu\nu} \omega^{\nu}$.
Indeed, we find
\begin{align}
\{ S_{\mu\nu} \omega^{\nu}, \tilde{\Lambda}_i{}^{\rho} \} = 0 ,
\end{align}
where we made use of
$\omega^{\mu} p_{\mu} / p = \omega_{\mu} \Lambda_{0}{}^{\mu} = \frac{1}{2} \omega_{\mu} \omega^{\mu}$
and $\tilde{\Lambda}_i{}^{\mu} w_{\mu} = \tilde{\Lambda}_i{}^{\mu} p_{\mu}$.
We have discovered the spin gauge constraint
\begin{equation}\label{SGC}
0 \approx \mathcal{C}_{\mu} := S_{\mu\nu} \left(\frac{p^{\nu}}{p} + \Lambda_{0}{}^{\nu}\right)
        \equiv S_{\mu\nu} \omega^{\nu} ,
\end{equation}
where weak equality \cite{Dirac:1964} (restriction to the constraint surface)
is denoted by $\approx$.
It is further illustrated below that this constraint physically makes sense.
This spin gauge constraint is not an SSC (in the usual sense)
and does not correspond to a choice for the worldline,
but it parametrizes the possible SSCs through a gauge field $\Lambda_{0}{}^{\nu}$, see
below for discussion.

We notice that Ref.~\cite{Hanson:Regge:1974} was missing the $\Lambda_{0}{}^{\nu}$
in this constraint.
One might object that the appearance of $\Lambda_{0}{}^{\nu}$
is breaking the SO(1,3) Lorentz invariance in its first index. However, this
index belongs to the body-fixed frame, whose time-like component is not
an observable. Indeed, we already argued that $\Lambda_{0}{}^{\nu}$ is
a physically irrelevant gauge degree of freedom.
Furthermore, also from an EFT point of view, only SO(3) rotation invariance
is a symmetry that must be respected for the body-fixed frame
\cite{Goldberger:2007,Goldberger:Rothstein:2006}, instead of SO(1,3) Lorentz invariance.

We must have another constraint related to a gauge symmetry here, namely
that of reparametrization invariance of the worldline parameter $\sigma$.
The associated constraint is the mass shell one \cite{Hanson:Regge:1974}
\begin{equation}
0 \approx \mathcal{H} := p_{\mu}p^{\mu} - \mathcal{M}^2 ,
\end{equation}
which together with (\ref{SGC}) forms the basis for an action principle.

\section{Action for spinning particles}
We are going to construct an action principle based on a Hamiltonian.
The Dirac Hamiltonian $H_{D}$ \cite{Rosenfeld:1930, Dirac:1964, Pons:2005}
is the canonical Hamiltonian plus the
(primary) constraints, which are added with the help of
Lagrange multipliers. However, due to reparametrization invariance,
the canonical Hamiltonian vanishes, so that $H_{D}$ is just
composed of the constraints,
\begin{equation}
H_{D} = \frac{\lambda}{2} \mathcal{H} + \chi^{\mu} \mathcal{C}_{\mu} ,
\end{equation}
where $\lambda$ and $\chi^{\mu}$ are the Lagrange multipliers.

An action principle for spinning point-particles (PP) must
reproduce the equations of motion of $H_D$ with Poisson brackets
from Eq.~(\ref{poisson}).
\append{The relation between action and Poisson brackets is
discussed in Appendix \ref{PBfromaction}.}
The appropriate action reads \cite{Hanson:Regge:1974}
\begin{equation}\label{PPaction}
S_{\text{PP}} = \int d \sigma \left[ - p_{\mu} u^{\mu} - \frac{1}{2} S_{\mu\nu} \Omega^{\mu\nu} - H_{D} \right] ,
\end{equation}
where the variables with independent variations are
$z^{\mu}$, $p_{\mu}$, $\Lambda^{A\mu}$, and $S_{\mu\nu}$
and we introduced the abbreviations $u^{\mu} = \dot{z}^{\mu}$, 
$\Omega^{\mu\nu} = \Lambda_{A}{}^{\nu} \dot{\Lambda}^{A\mu}$,
and $\dot{~} = d / d \sigma$. Since $\Lambda^{A\mu}$ is a Lorentz
matrix, it can only be varied by an infinitesimal Lorentz
transformation $\delta \theta^{\mu\nu} = - \delta \theta^{\nu\mu}$, i.e.,
$\delta \Lambda^{A\mu} = \Lambda^{A\nu} \delta \theta_{\nu}{}^{\mu}$.


Notice that the dynamical mass $\mathcal{M}$ can be a general function of
the dynamical variables. Then the mass $\mathcal{M}$ contains all the
interaction energies. It must be adapted such that the point-particle
provides an effective description for some extended body on macroscopic scale
(reducing the almost innumerable internal degrees of freedom of the body to the relevant ones).
The dynamics is then encoded through $\mathcal{M}$ and the action principle.
This makes the dynamical mass $\mathcal{M}$ analogous to a thermodynamic potential, like the
internal energy. It should be noted that EFTs are of comparable importance for both
particle and statistical physics. The analogy of $\mathcal{M}$
to a thermodynamic potential suggests to apply a construction
of $\mathcal{M}$ using symmetries and power counting arguments,
as usual in an EFT. For black holes, $\mathcal{M}$ as a function of the dynamical
variables is related to the famous laws of black hole dynamics
\cite{Misner:Thorne:Wheeler:1973}, see also \cite{Steinhoff:2014:2}
for discussion.

We require here that the constraints in $H_D$ are related to gauge symmetries.
Then the Lagrange multipliers are not fixed by requiring that the
constraints are preserved in time and represent the gauge freedom.
This implies that the Poisson brackets between all pairs of constraints vanish weakly here.
Constraints with this property are called first class \cite{Dirac:1964}.
See Ref.~\cite{Pons:2005} for further discussion of gauge symmetry
in constrained Hamiltonian dynamics.
In contrast to our requirement that all three independent components of the
spin constraints are first class, only one component of the spin constraint
used in \cite{Hanson:Regge:1974} is first class.
This renders the completion of the set of constraints in \cite{Hanson:Regge:1974}
rather unsatisfactory, because only first class constraints require a completion
through gauge fixing constraints.

\section{Spin gauge constraint}
The first main objective of the present {\paper} is to establish
$\mathcal{C}_{\mu}$ as the proposed spin gauge constraint.
We just argued that it should
be a first class constraint. Furthermore, it should be
the generator of a spin gauge transformation, i.e.,
a shift of the representative worldline $\Delta z^{\mu}(\sigma)$
together with the appropriate change in the spin,
\begin{equation}
\Delta S_{\mu\nu} \approx 2 p_{[\mu} \Delta z_{\nu]} , \label{Sgtrafo}
\end{equation}
This is in agreement with the physical picture in Fig.\ \ref{centerfig},
which refers to the rest-frame where $p^{\mu} = (m, 0)$.
These are the physical requirements we have on $\mathcal{C}_{\mu}$.

These requirements are indeed met for Eq.~(\ref{SGC}).
It is easy to see that $\mathcal{C}_{\mu}$ is first-class among itself,
\begin{equation}
\{ \mathcal{C}_{\mu}, \mathcal{C}_{\nu} \} = \frac{2}{p} p_{[\mu} \mathcal{C}_{\nu]} \approx 0 ,
\end{equation}
so it qualifies for a gauge constraint.
Then the generator of an infinitesimal spin gauge transformation is
$\epsilon^{\mu} \mathcal{C}_{\mu}$ and the transformations of the
fundamental variables read
\begin{align}
\Delta z^{\mu} &:= \{ \epsilon^{\alpha} \mathcal{C}_{\alpha}, z^{\mu} \}
        = \frac{1}{p} P^{\mu\nu} S_{\nu\alpha} \epsilon^{\alpha}, \\
\Delta p^{\mu} &:= \{ \epsilon^{\alpha} \mathcal{C}_{\alpha}, p^{\mu} \}
        = 0 , \label{ptrafo} \\
\Delta \Lambda_A{}^{\mu} &:= \{ \epsilon^{\alpha} \mathcal{C}_{\alpha}, \Lambda_A{}^{\mu} \}
        = 2 \epsilon^{[\mu} \omega^{\nu]} \Lambda_{A\nu} , \label{Ltrafo} \\
\Delta S_{\mu\nu} &:= \{ \epsilon^{\alpha} \mathcal{C}_{\alpha}, S_{\mu\nu} \}
        = 2 p_{[\mu} \Delta z_{\nu]} - 2 \epsilon_{[\mu} \mathcal{C}_{\nu]} , \label{Strafo}
\end{align}
where $P^{\mu\nu}$ is the projector onto the spatial hypersurface of the rest-frame,
\begin{equation}
P^{\mu\nu} := \eta^{\mu\nu} - \frac{p^{\mu} p^{\nu}}{p^2} .
\end{equation}
Notice that $\Delta z^{\mu}$ is a spatial vector in the rest-frame of the
particle, $p_{\mu} \Delta z^{\mu} = 0$. This makes it obvious that the
symmetry group is three-dimensional.

Now, on the constraint surface, Eq.~(\ref{Strafo}) is identical
to our second and final requirement in Eq.~(\ref{Sgtrafo}).
That is, it is precisely the amount that an angular momentum changes
if the reference point is moved by $\Delta z^{\mu}$. This shows that
$\epsilon^{\mu} \mathcal{C}_{\mu}$ indeed generates a shift of the
reference worldline within the object and that $\mathcal{C}_{\mu}$
is the corresponding gauge constraint.

While the physical meaning of $\Delta S_{\mu\nu}$ is immediately
clear, an interpretation of $\Delta \Lambda_A{}^{\mu}$ deserves
a more detailed illustration. We recall that the physically relevant
components of $\Lambda_A{}^{\mu}$ are obtained by a finite boost to the
rest frame, see Eq.~(\ref{standardboost}). Therefore, $\Delta \Lambda_A{}^{\mu}$
should be given by a standard boost to the rest-frame followed by another
standard boost to a frame infinitesimally close to $\Lambda^{A\mu}$. It is
straightforward to check using the standard boost, Eq.~(\ref{standardboost}),
that this is the case.

\section{Spin gauge fixing}
As usual, a gauge fixing now requires a gauge condition, that is,
a condition on the gauge field $\Lambda_{0}{}^{\mu}$.
Notice that the spin gauge constraint (\ref{SGC}) does not correspond
to a choice for a representative worldline, because it contains
the unspecified gauge field $\Lambda_{0}{}^{\mu}$. 
The following choices turn the spin gauge constraint (\ref{SGC}) into
familiar choices for the SSC:
\begin{align}
\Lambda_{0}{}^{\mu} &\approx \frac{p^{\mu}}{p} \quad \Rightarrow \quad
        \mathcal{C}_{\mu} = S_{\mu\nu} p^{\nu} \approx 0 , \\
\Lambda_{0}{}^{\mu} &\approx \delta_0^{\mu} \quad \Rightarrow \quad
        \mathcal{C}_{\mu} = S_{\mu\nu} ( p^{\nu} + p \delta_0^{\nu} ) \approx 0 , \\
\Lambda_{0}{}^{\mu} &\approx \frac{2 p^0 \delta_0^{\mu} - p^{\mu}}{p} \quad \Rightarrow \quad
        \mathcal{C}_{\mu} = S_{\mu0} \approx 0 .
\end{align}
The first one is due to Fokker \cite{Fokker:1929},
the second due to Pryce, Newton, and Wigner \cite{Pryce:1935, Pryce:1948, Newton:Wigner:1949},
and the last one due to Pryce and M{\o}ller \cite{Pryce:1948, Moller:1949}.
In general relativity, the first condition was first considered by
W.~M.~Tulczyjew \cite{Tulczyjew:1959}, the third one by
Corinaldesi and Papapetrou \cite{Corinaldesi:Papapetrou:1951},
and the second one was applied more recently only
\cite{Porto:Rothstein:2006, Levi:2008, Steinhoff:Schafer:Hergt:2008, Barausse:Racine:Buonanno:2009}
in slightly different forms and contexts. The differences arise in the
way the (normalized) time vector $\delta_0^{\mu}$ is generalized to curved spacetime.
However, it should be noted that \cite{Porto:Rothstein:2006, Levi:2008, Barausse:Racine:Buonanno:2009}
suggest to complete the set of constraints by $\Lambda_0{}^{\mu} \propto p^{\mu}$, while
here in the context of spin gauge symmetry this would lead to the first SSC, but not
to the second one. The SSC and the condition on $\Lambda_0{}^{\mu}$ can not
be chosen independently here.

Any of the above conditions can be added to the set of constraints, e.g.,
$0 \approx \Lambda_{0}{}^{\mu} - \delta_0^{\mu}$. This constraint then
turns the spin gauge constraint into a second class constraint, which means that
the set of constraints can be eliminated using the Dirac bracket \cite{Dirac:1964}
(and one can solve for the Lagrange multipliers).
This is the usual manner in which gauge fixing is handled in the context
of constrained Hamiltonian dynamics.

However, one can alternatively insert the solution to the set of constraints
into the action in a classical context. For the case of the Pryce-Newton-Wigner spin gauge,
it holds $\Lambda_0{}^{\mu} = \delta_0^{\mu}$ and $\Lambda_A{}^0 = \delta_A^0$,
so the temporal components drop out of the spin kinematic term in the action,
\begin{align}\label{NWkin}
\frac{1}{2} S_{\mu\nu} \Omega^{\mu\nu}
        = \frac{1}{2} S_{ij} \Omega^{ij} , \quad 
\Omega^{ij}
        = - \Lambda^{ki} \dot{\Lambda}^{kj} .
\end{align}
The kinematic term still has the same form as in Eq.~(\ref{PPaction}), but the
indices are 3-dimensional now and the SO(1,3) Lorentz matrix was reduced to a
SO(3) rotation matrix. Therefore, the standard so(1,3) Lie algebra Poisson
bracket for the spin, Eq.~(\ref{poisson}), must be replaced by a standard so(3) algebra for the
spatial components of the spin. However, for other gauge choices, the kinematic
term will not simplify this drastically and the reduced
Poisson bracket algebra will in general be more complicated.
This makes the Pryce-Newton-Wigner gauge probably the most useful one
if one aims at a reduction of variables, while the Fokker-Tulczyjew one leads
to manifestly covariant equations of motion. It should be emphasized
that all gauges lead to equivalent equations of motion by construction here, if
the mass shell constraint $\mathcal{H}$ is spin gauge invariant.

Note that we obtain the same set of constraints as \cite{Hanson:Regge:1974}
if we choose the first gauge. However, adding a gauge symmetry to the
theory should not be seen as adding unnecessary complications. On the contrary, different
gauges are useful for different applications. For instance, in Ref.~\cite{Hanson:Regge:1974}
a transformation from Fokker-Tulczyjew to Pryce-Newton-Wigner variables is considered
because it simplifies the reduced Poisson brackets. Here one can directly
use the second gauge fixing condition instead and the simplification of
brackets is explained by Eq.~(\ref{NWkin}).
This is an important consequence of our construction:
Different SSCs are manifestly equivalent, in particular
the covariant SSC is equivalent to noncovariant ones by construction.

\section{Gauge invariant variables}
Now we turn to the second main objective of the present {\paper},
which is the construction of the mass shell constraint
$\mathcal{H}$ such that it is invariant under spin gauge
transformations. This is equivalent to
\begin{equation}
\{ \mathcal{C}_{\mu}, \mathcal{H} \} \approx 0 .
\end{equation}
The usual way to construct invariant quantities is by combining
objects with are invariant. That is, we aim
to find a position, spin, and Lorentz matrix which have weakly vanishing
Poisson bracket with $\mathcal{C}_{\mu}$. We already encountered the
invariant Lorentz matrix $\tilde \Lambda_i{}^{\mu}$ given by
Eq.~(\ref{linv}). Recalling that $\tilde \Lambda_i{}^{\mu}$ was
obtained by a boost to the rest frame,
we may guess that the following projections to the rest-frame variables,
\begin{align}
\tilde z^{\mu} := z^{\mu} + S^{\mu\nu} \frac{p_{\nu}}{p^2} , \label{invpos} \quad
\tilde S_{\mu\nu} := P_{\mu}{}^{\alpha} P_{\nu}{}^{\beta} S_{\alpha\beta}
\end{align}
have the desired properties. It is indeed straightforward to show that
these variables with a tilde have weakly vanishing
Poisson brackets with $\mathcal{C}_{\mu}$,
\begin{align}
\{ \mathcal{C}_{\alpha}, \tilde{z}^{\mu} \}
        &= - \frac{1}{p^2} [ \delta^{\mu}_{\alpha} p_{\nu}
                - \delta^{\mu}_{\nu} p_{\alpha} ] \mathcal{C}^{\nu}
        \approx 0 , \\
\{ \mathcal{C}_{\alpha}, \tilde{S}_{\mu\nu} \}
        &= - 2 P_{\alpha[\mu} P_{\nu]}{}^{\beta} \mathcal{C}_{\beta}
        \approx 0 ,
\end{align}
and the linear momentum is already invariant, see Eq.~(\ref{ptrafo}).
Therefore, if the mass shell constraint depends on these variables only, then
it is invariant under spin gauge transformation,
\begin{equation}
\mathcal{M} = \mathcal{M}(\tilde{z}^{\mu}, p^{\mu},
        \tilde{\Lambda}_i{}^{\mu}, \tilde{S}_{\mu\nu})
\quad \Rightarrow \quad
        \{ \mathcal{C}_{\mu}, \mathcal{H} \} \approx 0 .
\end{equation}

The most simple model is given by
\begin{equation}\label{trajectory}
\mathcal{M}^2 = f(\tilde{S}^2)
\quad \text{in} \quad
\mathcal{H} = p_{\mu} p^{\mu} - \mathcal{M}^2 ,
\end{equation}
and $\tilde{S}^2 = \frac{1}{2} \tilde{S}_{\mu\nu} \tilde{S}^{\mu\nu}$.
Here $f$ is an almost arbitrary function, commonly referred to as a Regge trajectory
\cite{Hanson:Regge:1974}. We require it to be analytic and nonconstant
(for $f=\text{const}$ it follows $\Omega^{\mu\nu}=0$ for any spin).
This function encodes the rotational kinetic energy and the moment of inertia
of the body. For black holes, it is related to the laws of black hole mechanics.

\section{Simplified invariant action}
The construction of the last section has a problematic aspect.
All fields interactions, entering through
the dynamical mass $\mathcal{M}$, must be taken at the position $\tilde{z}^{\mu}$,
which in general is different from the worldline coordinate $z^{\mu}$. This
is particularly a problem for coupling to the gravitational field, because
first of all the tangent spaces at $\tilde{z}^{\mu}$ and $z^{\mu}$ are different and
second the difference between the positions is not a tangent vector. Therefore,
it is not straightforward to generalize the current construction to the
general relativistic case, where the dynamical mass must be coordinate invariant.

However, because the mass shell constraint must be written entirely in terms
of the position $\tilde{z}^{\mu}$, it is suggestive to shift the worldine of the
action to this position. It should be noted that one can not switch to the invariant spin
and Lorentz matrix as fundamental variables, because the transformation
involves projections (notice $\tilde{\Lambda}_i{}^{\mu} p_{\mu} = 0$). This leads to an action containing a time derivative of the momentum,
\begin{equation}\label{invaction}
S_{\text{PP}} = \int d \sigma \left[ - p_{\mu} \tilde{u}^{\mu}
	- S^{\mu\nu} \frac{\dot{p}_{\mu} p_{\nu}}{p^2}
	- \frac{1}{2} S_{\mu\nu} \Omega^{\mu\nu} - H_{D} \right] .
\end{equation}
The Poisson brackets involving $\tilde{z}^{\mu}$ will not be standard canonical,
but they can be readily obtained from Eq.~(\ref{invpos}) and the old Poisson brackets.
Note that one can associate Poisson brackets to an action
if it contains at most first order time derivatives (and no pathologies arise).
\append{See the Appendix~\ref{PBfromaction}.}
The equations of motion are still first order, which is
important because otherwise more initial values would be needed.

Now it is now straightforward to couple
the spinning particle to the gravitational field, namely by replacing ordinary
derivatives with respect to $\sigma$ in Eq.~(\ref{invaction}) by covariant ones (minimal coupling).
Additionally, nonminimal couplings can be added in the sense of an
effective theory via the dynamical mass $\mathcal{M}$, as long as these are evaluated
at the position of the new worldline $\tilde{z}^{\mu}$ and constructed using the matter variables
$\tilde{\Lambda}_i{}^{\mu}$, $\tilde{S}_{\mu\nu}$, and $p_{\mu}$.
Further development of the general relativistic case and an
application to the post-Newtonian approximation is given in
\cite{Levi:Steinhoff:2014:3}.
In the following section, we illustrate that this construction
is also convenient for coupling to other fields, like the
electromagnetic one, and see that nonminimal couplings represent
the multipoles of the body \cite{Bailey:Israel:1975,
Porto:Rothstein:2008:2, Levi:Steinhoff:2014:2}.

The time derivative of the momentum in our action is similar to
the acceleration term introduced in \cite{Yee:Bander:1993}.
(It also appeared in a similar time+space decomposed
form in \cite{Hergt:Steinhoff:Schafer:2011, Levi:Steinhoff:2014:1}.)
After a coupling to gravity (or the electromagnetic field), one can
approximately remove this term, if this is desired, using
a manifestly covariant shift of the worldline as introduced in
\cite{Bailey:Israel:1980}. This approximately corresponds to inserting
the equation of motion for $\dot{p}_{\mu}$ into the action \cite{Damour:Schafer:1991}.
This transformation leads to the nonminimal coupling proportional
to the Fokker-Tulczyjew SSC used in \cite{Porto:Rothstein:2008:2}. However, in
both Refs.~\cite{Yee:Bander:1993, Porto:Rothstein:2008:2} this term was introduced in order to
preserve the covariant SSC, while here it arises from the requirement
of spin gauge symmetry. This distinction is significant in the context of
an EFT.

\section{Electromagnetic interaction}
A minimal coupling to the electromagnetic field with charge $q$ can be
introduced as usual by adding $- q \tilde{A}_{\mu} \tilde{u}^{\mu}$ to the Lagrangian,
see, e.g., Eq.~(6.1) in \cite{Arnowitt:Deser:Misner:1962, Arnowitt:Deser:Misner:2008},
\begin{align}\label{emaction}
S_{\text{PP}} &= \int d \sigma \bigg[ - p_{\mu} \tilde{u}^{\mu}
	- q \tilde{A}_{\mu} \tilde{u}^{\mu} \nonumber \\
&\quad	- S^{\mu\nu} \frac{\dot{p}_{\mu} p_{\nu}}{p^2}
	- \frac{1}{2} S_{\mu\nu} \Omega^{\mu\nu} - H_{D} \bigg] ,
\end{align}
where $\tilde{A}_{\mu} = A_{\mu}(\tilde{z}^{\nu})$.
This new term turns into a total time derivative under electromagnetic gauge
transformations.
The added term is also manifestly invariant under spin gauge transformations,
because it only involves invariant quantities. Here it is convenient that
we shifted the worldline to the spin-gauge-invariant position $\tilde{z}^{\mu}$.

We are going to derive the equations of motion belonging to the
action (\ref{emaction}) and explicitly construct the Poisson brackets associated
to it. The $\delta p_{\mu}$-variation leads to the velocity-momentum relation
\begin{equation}
\begin{split}\label{uEOM}
\tilde{u}^{\mu} &=
	2 \tilde{S}^{\mu\nu} \frac{\dot{p}_{\nu}}{p^2}
	+ \dot{S}^{\mu\nu} \frac{p_{\nu}}{p^2}
	- \frac{\partial H_D}{\partial p_{\mu}} ,
\end{split}
\end{equation}
and from the $\delta S_{\mu\nu}$-variation it follows
\begin{equation}
\begin{split}
\Omega^{\mu\nu} &= \frac{2 p^{[\mu} \dot{p}^{\nu]}}{p^2}	
	- 2 \frac{\partial H_D}{\partial S_{\mu\nu}} .
\end{split}
\end{equation}
Finally, the $\delta \tilde{z}^{\mu}$-variation leads to
\begin{equation}
\begin{split}
\dot{p}_{\mu} = - q \tilde{F}_{\mu\nu} \tilde{u}^{\nu} + \frac{\partial H_D}{\partial \hat{z}^{\mu}} ,
\end{split}
\end{equation}
with the Faraday tensor $\tilde{F}_{\mu\nu} := \tilde{A}_{\mu,\nu} - \tilde{A}_{\nu,\mu}$,
and the $\delta \Lambda^{A\mu}$-variation gives
\begin{equation}
\begin{split}
\dot{S}_{\mu\nu} = - 2 \Omega^{\rho}{}_{[\mu} S_{\nu]\rho}
	+ 2 \Lambda_{A[\mu} \frac{\partial H_D}{\partial \Lambda_A{}^{\nu]}} ,
\end{split}
\end{equation}
We eliminate the time derivatives on the right hand side of Eq.~(\ref{uEOM}),
\begin{align}
\tilde{u}^{\mu} &=
	- \frac{q}{p^2} \tilde{S}^{\mu\nu} \tilde{F}_{\nu\alpha} \tilde{u}^{\alpha}
	+ U^{\mu} , \\
\begin{split}
U^{\mu} &:= - \frac{\partial H_D}{\partial p_{\mu}}
	- 4 \delta_{\alpha}^{[\mu} S^{\nu]}{}_{\beta} \frac{p_{\nu}}{p^2} \frac{\partial H_D}{\partial S_{\alpha\beta}} \nl
	+ \frac{\tilde{S}^{\mu\nu}}{p^2} \frac{\partial H_D}{\partial \hat{z}^{\nu}}
	+ 2 \frac{\eta^{\mu[\alpha} p^{\nu]}}{p^2} \Lambda_{A\alpha} \frac{\partial H_D}{\partial \Lambda_A{}^{\nu}} ,
\end{split}
\end{align}
which can be solved for $\tilde{u}^{\mu}$ using Eq.~(A.4) and (A.7)
in Ref.~\cite{Hanson:Regge:1974},
\begin{equation}
\tilde{u}^{\mu} = \left[ \delta_{\alpha}^{\mu} - \frac{q \tilde{S}^{\mu\nu} \tilde{F}_{\nu\alpha}}{p^2 - \frac{1}{2} q \tilde{S}^{\rho\delta} \tilde{F}_{\rho\delta} } \right] U^{\alpha} ,
\end{equation}
where we used $\tilde{S}_{\mu\nu} *\!\tilde{S}^{\mu\nu} = 0$ ($*$ denotes the Hodge dual).
Using this relation, all time derivatives can be removed from the
right hand sides of the equations of motion. The equation of motion of
a dynamical variable is then expressed in terms of derivatives of $H_D$ with respect to
other dynamical variables only, and the prefactor is the mutual Poisson bracket.
\append{Explicit expressions for all Poisson brackets are given in Appendix \ref{PBfromaction}.}
These Poisson brackets are similar to Eq.~(5.29) in \cite{Hanson:Regge:1974}.
However, for most applications it should be sufficient to have
an action principle and the equations of motion in the form given above.
But it is good to know that the equations of motion follow a
symplectic flow and that the Poisson brackets can be obtained explicitly if needed.
An analogous calculation should be possible for the gravitational interaction.

The finite-size and internal structure of the particle is modelled
by nonminimal couplings in the dynamical mass $\mathcal{M}$.
These are composed of the (electromagnetic gauge invariant) Faraday tensor $F_{\mu\nu}$
and its derivatives, where an increasing number of derivatives corresponds to smaller
length scales or higher multipoles.
As an illustration for the treatment of electromagnetic
multipoles, we consider the simplest case of a dipole, which corresponds to
a nonminimal coupling to the Faraday tensor $F_{\mu\nu}$ in the dynamical mass.
For a spin-induced dipole, this reads
\begin{equation}
\mathcal{M}^2 = f(\tilde{S}^2) + \frac{g q}{2} \tilde{F}_{\mu\nu} \tilde{S}^{\mu\nu} ,
\end{equation}
where $g$ is the gyromagnetic ratio.
For this model, our equations of motion are in agreement
with \cite{Hanson:Regge:1974}. Making the gauge
choice $\Lambda_0{}^{\mu} = p^{\mu} / p$ and requiring that it is preserved in
time by above equations of motions, we find that $\chi^{\mu} = 0$ for the
Lagrange multiplier of the spin gauge constraint in this case. A contraction of (\ref{uEOM}) with
$p_{\mu}$ then leads to $\lambda = \tilde{u}^{\mu} p_{\mu} / p^2$ for the
Lagrange multiplier of the mass shell constraint, which is related to the
parametrization of the worldline. Then we find structural agreement with
(4.29), (4.33), and (4.34) in \cite{Hanson:Regge:1974} and therefore also with
the Bargmann-Michel-Telegdi equations \cite{Bargmann:Michel:Telegdi:1959}. The normalization
of the dipole interaction differs compared to \cite{Hanson:Regge:1974},
which is due to the fact that in \cite{Hanson:Regge:1974} the dipole is
proportional to the angular velocity, while here it is proportional to the spin.
In \cite{Bargmann:Michel:Telegdi:1959} the SSC $S^{\mu\nu} u_{\mu} = 0$
\cite{Frenkel:1926, Mathisson:1937, Mathisson:2010}
is used, which fails to uniquely
define a worldline and in general leads to helical motion
\cite{Mathisson:1937, Mathisson:2010, Costa:etal:2011}.
Finally, we notice that a dipole linearly induced by an external field is
modelled by couplings of the form shown in Eq.~(1) of Ref.~\cite{Goldberger:Rothstein:2006:2},
which can be added to the dynamical mass here.

If desired, one can shift the $\dot{p}_{\mu}$-term in the action (\ref{emaction})
to higher orders in the derivative of the electromagnetic field
through redefinitions of the variables \cite{Damour:Schafer:1991}.
That is, through successive redefinitions, one can turn the $\dot{p}_{\mu}$-term
into nonminimal interaction terms of increasing derivative order.
Consistent with neglecting finite-size effects of a certain
multipolar order in $\mathcal{M}$, one can terminate this
process at the desired order and the $\dot{p}_{\mu}$-term is
effectively removed. The physical reason for the additional nonminimal interactions
is that a shift of position of a monopole $q$
generates an infinite series of higher multipoles \cite{Bailey:Israel:1980}.
\remark{Is it possible to do this in closed form?}

\section{Conclusions}
The spin gauge symmetry formulated in the present {\paper} is an
important ingredient to extend the EFT for charged particles to spinning charged particles
(see, e.g., \cite{Galley:Leibovich:Rothstein:2010}).
This should serve as a classical EFT for massive fermions.
Massless particles have a different little group, so our approach
need several adjustments for this case.
We identified spin gauge invariant variables in the present {\paper}, which should be useful
for a matching of the EFT.

The general relativistic case is analogous to the electromagnetic one
and is invaluable for modelling the motion of black holes and neutron stars.
This will lead to better gravitational wave forms needed for the data analysis
requirements of future gravitational wave astronomy.

\acknowledgments
We acknowledge inspiring discussions with Gerhard Sch{\"a}fer and Michele Levi.
A part of this work was supported by FCT (Portugal) through grants
SFRH/BI/52132/2013 and PCOFUND-GA-2009-246542 (cofunded by Marie Curie Actions).

\append{
\appendix

\section{Poisson brackets from the action\label{PBfromaction}}
Consider an action containing at most first order derivatives in time.
We can write it in the form 
\begin{equation}
S = \int dt \left[ B_a(q^b) \, \dot{q}^a - H(q^b) \right] ,
\end{equation}
where $a$, $b$ label the dynamical variables $q_a$.
The equations of motion read
\begin{equation}
M_{ab} \dot{q}^b = \partial_a H , \quad
M_{ab} := \partial_a B_b - \partial_b B_a .
\end{equation}
where $\partial_a = \partial / \partial q^a$, or
\begin{equation}\label{genEOM}
\dot{q}^a = M^{ab} \partial_b H , \quad
M^{ab} := M_{ab}^{-1}
\end{equation}
This can be written using Poisson brackets
\begin{equation}\label{defPB}
\dot{q}^a = \{ H, q^a \} , 
\end{equation}
if we set
\begin{equation}
\{ X, Y \} = M^{ba} \partial_a X \partial_b Y .
\end{equation}

Instead of computing $M_{ab}$ and its inverse directly from the action,
it is often easier to obtain the equations of motion and
transform them to the form of Eq.~(\ref{genEOM}).
Then one can directly read off $M_{ab}^{-1}$.
For the case of electromagnetically interacting spinning particles (\ref{emaction}), we have
$\tilde{u}^{\mu} = G^{\mu}{}_{\nu} U^{\nu}$ where
\begin{equation}
G^{\mu}{}_{\nu} = \delta_{\nu}^{\mu} - \frac{q}{P^2} \tilde{S}^{\mu\alpha} \tilde{F}_{\alpha\nu} , \quad
P^2 = p^2 - \frac{1}{2} q \tilde{S}^{\mu\nu} \tilde{F}_{\mu\nu} .
\end{equation}
This leads to the Poisson brackets involving $\tilde{z}^{\mu}$
\begin{align}
\{ \hat{z}^{\mu}, \tilde{z}^{\nu} \} &= G^{\nu}{}_{\alpha} \frac{\tilde{S}^{\alpha\mu}}{p^2}
	= - \frac{\tilde{S}^{\mu\nu}}{P^2} , \\
\{ p_{\mu}, \tilde{z}^{\nu} \} &= - G^{\nu}{}_{\mu} , \\
\{ \Lambda_A{}^{\mu}, \tilde{z}^{\nu} \} &= \frac{2}{p^2} G^{\nu[\alpha} p^{\mu]} \Lambda_{A\alpha} , \\
\{ S_{\mu\nu}, \tilde{z}^{\alpha} \} &= - \frac{4}{p^2} G^{\alpha}{}_{\beta} \delta^{[\beta}_{[\mu} S^{\rho]}{}_{\nu]} p_{\rho} ,
\end{align}
where we used Eq.~(A.5) in \cite{Hanson:Regge:1974},
\begin{equation}
G^{\nu}{}_{\alpha} \frac{\tilde{S}^{\alpha\mu}}{p^2}
	= - \frac{\tilde{S}^{\mu\nu}}{P^2}
	= - G^{\mu}{}_{\alpha} \frac{\tilde{S}^{\alpha\nu}}{p^2} .
\end{equation}
Similarly, from considering the equation of motion for $p_{\mu}$,
\begin{align}
\{ p_{\mu}, p_{\nu} \} &= q \tilde{F}_{\nu\alpha} G^{\alpha}{}_{\mu} , \\
\{ \Lambda_A{}^{\mu}, p_{\nu} \} &= - \frac{2 q}{p^2} \tilde{F}_{\nu\beta} G^{\beta[\alpha} p^{\mu]} \Lambda_{A\alpha} , \\
\{ S_{\mu\nu}, p_{\alpha} \} &= \frac{4 q}{p^2} \tilde{F}_{\alpha\delta} G^{\delta}{}_{\beta} \delta^{[\beta}_{[\mu} S^{\rho]}{}_{\nu]} p_{\rho} ,
\end{align}
and from the remaining equations
\begin{align}
\{ \Lambda_A{}^{\mu}, \Lambda_B{}^{\nu} \} &= - \frac{4 q}{p^4} \Lambda_{B\rho} p^{[\rho} \tilde{F}^{\nu]}{}_{\beta} G^{\beta[\alpha} p^{\mu]} \Lambda_{A\alpha} , \\
\begin{split}
\{ S_{\mu\nu}, \Lambda_A{}^{\alpha} \} &= - 2 \Lambda_A{}^{\beta} \eta_{\beta[\mu} \delta^{\alpha}_{\nu]} \nl
	+ \frac{8 q}{p^4} \Lambda_{A\delta} p^{[\delta} \tilde{F}^{\alpha]}{}_{\sigma} 
	G^{\sigma}{}_{\beta} \delta^{[\beta}_{[\mu} S^{\rho]}{}_{\nu]} p_{\rho} ,
\end{split}\\
\begin{split}
\{ S_{\mu\nu}, S_{\alpha\beta} \} &= - 4 S_{\alpha][\mu} \eta_{\nu][\beta} \nl
	+ \frac{16 q}{p^4} S^{\gamma}{}_{[\alpha} \delta_{\beta]}^{\delta} p_{[\delta} \tilde{F}_{\gamma]\sigma} 
	G^{\sigma}{}_{\chi} \delta^{[\chi}_{[\mu} S^{\rho]}{}_{\nu]} p_{\rho} .
\end{split}
\end{align}
The Poisson brackets are rather complicated. For most applications, it
is therefore better to work directly with the action (\ref{emaction})
if possible.
\remark{TODO: As a consistency check, one can confirm that $\mathcal{C}_{\mu}$
is first class using these brackets. This must hold by construction, since
the minimal coupling term $- q \tilde{A}_{\mu} \tilde{u}^{\mu}$ is spin
gauge invariant. Alternatively, one can check that the Lagrange multipliers
can not be fixed using the equations of motions.}

\section{Another check against Ref.~\cite{Hanson:Regge:1974}}
The term involving $\dot{p}_{\mu}$ in Eq.~(\ref{emaction}) can be removed by variable
redefinitions, which will then be canonical variables because the action
assumes a canonical form. Here we only intend to make a connection to the
results in \cite{Hanson:Regge:1974}. Therefore we apply the gauge
fixing $\Lambda_0{}^{\mu} = p^{\mu} / p$, or the SSC $S_{\mu\nu} p^{\mu} = 0$.
The term containing $\dot{p}_{\mu}$ can be cancelled from the
action by shifting back to the position $z^{\mu}$. Now new contributions
arise from the minimal coupling term. However, at the level of the action,
we can neglect terms of quadratic or higher order in the SSC $S_{\mu\nu} p^{\mu}$.
Then one can absorb all additional terms by defining the canonical momentum
\begin{equation}
	p_{\mu}^{\text{can}} = p_{\mu} + q A_{\mu} + q F_{\mu\nu} S^{\nu\rho} \frac{p_{\rho}}{p^2} .
\end{equation}
This can be solved for $p_{\mu}$ using the identities in Appendix A of
\cite{Hanson:Regge:1974} and one finds agreement of (\ref{trajectory}) with
(5.18) in \cite{Hanson:Regge:1974} (therein it
holds $\pi_{\mu} = p_{\mu}^{\text{can}} - q A_{\mu}$).
}

\ifnotprd
\bibliographystyle{utphys}
\fi

\ifarxiv
\providecommand{\href}[2]{#2}\begingroup\raggedright\endgroup

\else
\bibliography{../references,spingauge}
\fi

\end{document}